\documentclass[11pt,a4paper]{article}
\usepackage[hyperref]{acl2021}
\usepackage{times}
\usepackage{latexsym}

\usepackage{amsmath,amssymb}
\usepackage{booktabs}
\usepackage{tikz}
\usepackage{tipa}

\usepackage{pgfplots}

\usepackage{microtype}

\aclfinalcopy 
\def\aclpaperid{2659} 

\newcommand{\norm}[1]{\left\lVert#1\right\rVert}

\title{An Improved Model for Voicing Silent Speech}

\author{David Gaddy \and Dan Klein \\
  University of California, Berkeley \\
  \texttt{\{dgaddy,klein\}@berkeley.edu}}

\date{}

\begin{document}
\maketitle
\begin{abstract}
In this paper, we present an improved model for voicing silent speech, where audio is synthesized from facial electromyography (EMG) signals.  To give our model greater flexibility to learn its own input features, we directly use EMG signals as input in the place of hand-designed features used by prior work.  Our model uses convolutional layers to extract features from the signals and Transformer layers to propagate information across longer distances.  To provide better signal for learning, we also introduce an auxiliary task of predicting phoneme labels in addition to predicting speech audio features.  On an open vocabulary intelligibility evaluation, our model improves the state of the art for this task by an absolute 25.8\%.
\end{abstract}

\section{Introduction}

EMG-based voicing of silent speech is a task that aims to synthesize vocal audio from muscular signals captured by electrodes on the face while words are silently mouthed \cite{gaddy-klein-2020-digital,Toth2009Synthesizing}.
While recent work has demonstrated a high intelligibility of generated audio when restricted to a narrow vocabulary \cite{gaddy-klein-2020-digital}, in a more challenging open vocabulary setting the intelligibility remained low (68\% WER).
In this work, we introduce an new model for voicing silent speech that greatly improves intelligibility.

We achieve our improvements by modifying several different components of the model.
First, we improve the input representation.  While prior work on EMG speech processing uses hand-designed features \cite{jou2006towards,diener2015direct,meltzner2018development,gaddy-klein-2020-digital} which may throw away some information from the raw signals, our model learns directly from the complete signals 
with minimal pre-processing by using a set of convolutional neural network layers as feature extractors.
This modification follows recent work in speech processing from raw waveforms \cite{collobert2016wav2letter,schneider2019wav2vec} and gives our model the ability to learn its own features for EMG.

Second, we improve the neural architecture of the model.  While other silent speech models have been based around recurrent layers such as LSTMs \cite{janke2017emg,gaddy-klein-2020-digital}, we use the self-attention-based Transformer architecture \cite{vaswani2017attention}, which has been shown to be a more powerful replacement across a range of tasks.

Finally, we improve the signal used for learning.  Since the relatively small data sizes for this task creates a challenging learning problem, we introduce an auxiliary task of predicting phoneme labels to provide additional guidance.
This auxiliary loss follows prior work on related tasks of generating speech from ultrasound and ECoG sensors that also benefited from prediction of phonemic information \cite{toth2018multi,anumanchipalli2019speech}.

We evaluate intelligibility of audio synthesized by our model on the single-speaker data from \citet{gaddy-klein-2020-digital} in the most challenging open-vocabulary setting.
Our results reflect an absolute improvement in error rate of 25.8\% over the state of the art, from 68.0\% to 42.2\%, as measured by automatic transcription.  Evaluation by human transcription gives an even lower error rate of 32\%.

\section{Model}

At a high level, our system works by predicting a sequence of speech features from EMG signals and using a WaveNet vocoder \cite{Oord2016WaveNet} to synthesize audio from those predicted features, as was done in \citet{gaddy-klein-2020-digital}.
The first component, dubbed the transduction model, takes in EMG signals from eight electrodes around the face and outputs a sequence of speech features represented as Mel-frequency cepstral coefficients (MFCCs).
The final step of vocoding audio from MFCCs is unchanged in our work, so we defer to \citet{gaddy-klein-2020-digital} for the details of the WaveNet model.

The neural architecture for our transduction model is made up of a set of residual convolution blocks followed by a transformer with relative position embeddings, as shown in Figure~\ref{fig:overview}.
We describe these two components in Sections~\ref{sec:conv} and \ref{sec:transformer} below.
Next, in Section~\ref{sec:training} we describe our training procedure, which aligns each silent utterance to a corresponding vocalized utterance as in \citet{gaddy-klein-2020-digital} but with some minor modifications.
Finally, in Section~\ref{sec:phoneme} we describe the auxiliary phoneme-prediction loss that provides additional signal to our model during training.\footnote{Code for our model is available at \url{https://github.com/dgaddy/silent_speech}.}

\begin{figure}
    \centering
    \includegraphics[width=.9\columnwidth]{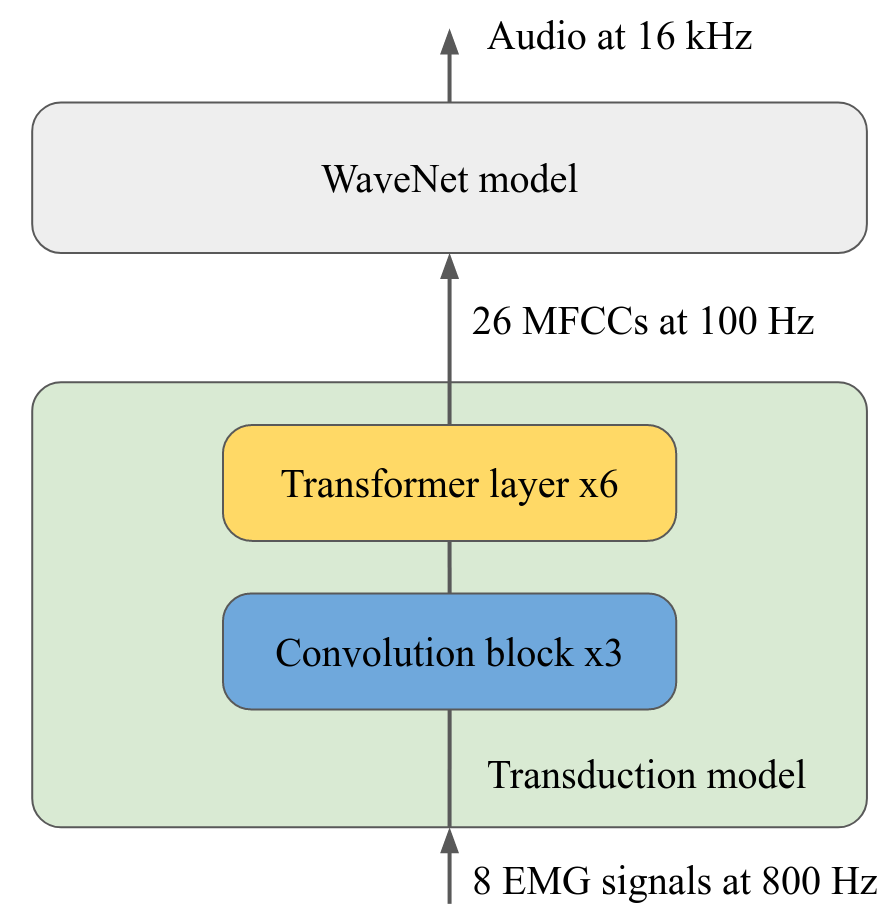}
    \caption{Model overview}
    \label{fig:overview}
\end{figure}

\subsection{Convolutional EMG Feature Extraction}
\label{sec:conv}

The convolutional layers of our model are designed to directly take in EMG signals with minimal preprocessing.
Prior to use of the input EMG signals, AC electrical noise is removed using band stop filters at harmonics of 60 Hz, and DC offset and drift are removed with a 2 Hz high-pass filter.
The signals are then resampled from 1000 Hz to 800 Hz, and the magnitudes are scaled down by a factor of 10.

\begin{figure}
    \centering
    \includegraphics[width=.95\columnwidth]{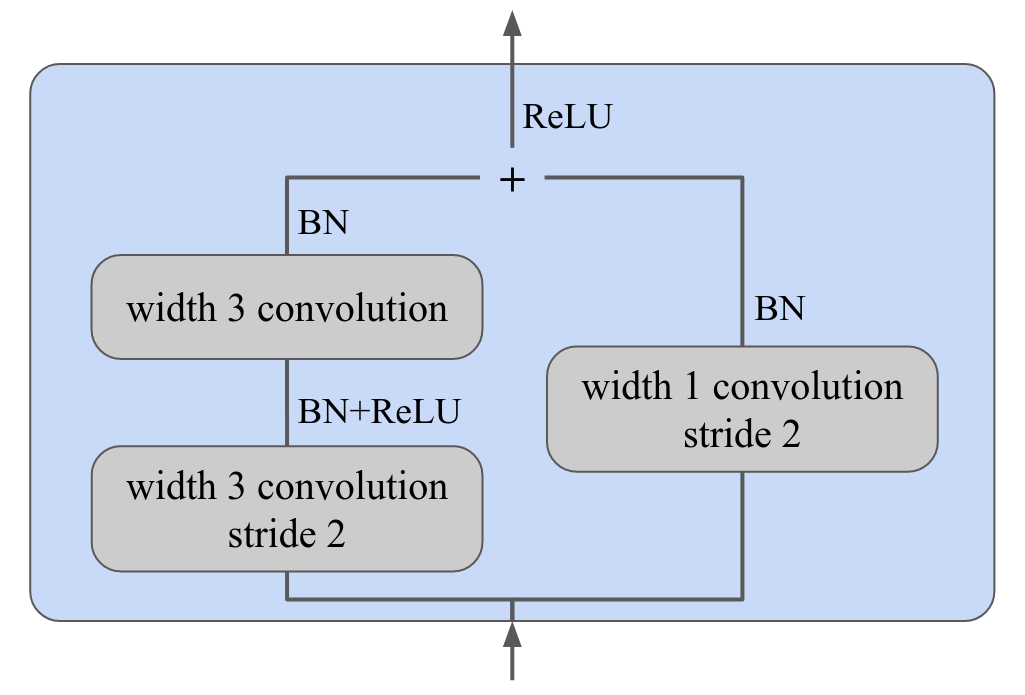}
    \caption{Convolution block architecture}
    \label{fig:conv-block}
\end{figure}

Our convolutional architecture uses a stack of 3 residual convolution blocks inspired by ResNet \cite{he2016deep}, but modified to use 1-dimensional convolutions.
The architecture used for each convolution block is shown in Figure~\ref{fig:conv-block}, and has two convolution-over-time layers along the main path as well as a shortcut path that does not do any aggregation over the time dimension.
Each block downsamples the signal by a factor of 2, so that the input signals at 800 Hz are eventually transformed into features at 100Hz to match the target speech feature frame rate.
All convolutions have channel dimension 768.

Before passing the convolution layer outputs to the rest of the model, we include an embedding of the session index, which helps the model account for differences in electrode placement after electrodes are reattached for each session.
Each session is represented with a 32 dimensional embedding, which is projected up to 768 dimensions with a linear layer before adding to the convolution layer outputs at each timestep.

\subsection{Transformer with Relative Position Embeddings}
\label{sec:transformer}

To allow information to flow across longer time horizons, we use a set of bidirectional Transformer encoder layers \cite{vaswani2017attention} on top of the convolution layers in our model.
To capture the time-invariant nature of the task, we use relative position embeddings as described by \citet{shaw-etal-2018-self} rather than absolute position embeddings.
In this variant, a learned vector $p$ that depends on the relative distance between the query and key positions is added to the key vectors when computing attention weights.  Thus, the attention logits are computed with
$$e_{ij}=\frac{(W_Kx_j+p_{ij})^\top(W_Qx_i)}{\sqrt{d}}$$
where $p_{ij}$ is an embedding lookup with index $i-j$, up to a maximum distance $k$ in each direction ($x$ are inputs to the attention module, $W_Q$ and $W_K$ are query and key transformations, and $d$ is the dimension of the projected vectors $W_Qx_i$).
For our model, we use $k=100$ (giving each layer a 1 second view in each direction) and set all attention weights with distance greater than $k$ to zero.
We use six of these Transformer layers, with 8 heads, model dimension 768, feedforward dimension 3072, and dropout 0.1.

The output of the last Transformer layer is passed through a final linear projection down to 26 dimensions to give the MFCC audio feature predictions output by the model.

\subsection{Alignment and Training}
\label{sec:training}

Since silent EMG signals and vocalized audio features must be recorded separately and so are not time-aligned, we must form an alignment between the two recordings to calculate a loss on predictions from silent EMG.
Our alignment procedure is similar to the predicted-audio loss used in \citet{gaddy-klein-2020-digital}, but with some minor aspects improved.

Our loss calculation takes in a sequence of MFCC features $\hat{A}_S$ predicted from silent EMG and another sequence of target features $A_V$ from a recording of vocalized audio for the same utterance.  We compute a pairwise distance between all pairs of features $$\delta[i,j]=\norm{A_V[i]-\hat{A}_S[j]}_2$$ and run dynamic time warping \cite{rabiner1993} to find a minimum-cost monotonic alignment path through the $\delta$ matrix.
We represent the alignment as $a[i]\rightarrow j$ with a single position $j$ in $\hat{A}_S$ for every index $i$ in $A_V$, and take the first such position when multiple are given by dynamic time warping.
The loss is then the mean of aligned pairwise distances:
$$L=\frac{1}{N_V}\sum_{i=1}^{N_V} \delta[i,a[i]]$$

In addition to the silent-EMG training, we also make use of EMG recordings during vocalized speech which are included in the data from \citet{gaddy-klein-2020-digital}.
Since the EMG and audio targets are recorded simultaneously for these vocalized examples, we can calculate the pairwise distance loss directly without any dynamic time warping.  We train on the two speaking modes simultaneously.

To perform batching across sequences of different lengths during training, we concatenate a batch of EMG signals across time then reshape to a batch of fixed-length sequences before feeding into the network.
Thus if the fixed batch-sequence-length is $l$, the sum of sample lengths across the batch is $N_S$, and the signal has $c$ channels, we reshape the inputs to size $(\left\lceil N_S/l\right\rceil,l,c)$ after zero-padding the concatenated signal to a multiple of $l$.  After running the network to get predicted audio features, we do the reverse of this process to get a set of variable-length sequences to feed into the alignment and loss described above.
This batching strategy allows us to make efficient use of compute resources and also acts as a form of dropout regularization where slicing removes parts of the nearby input sequence.
We use a sequence length $l=1600$ (2 seconds) and select batches dynamically up to a total length of $N_{Smax}=204800$ samples (256 seconds).

We train our model for 80 epochs using the AdamW optimizer \cite{loshchilov2017decoupled}.  The peak learning rate is $10^{-3}$ with a linear warm-up of 500 batches, and the learning rate is decayed by half after 5 consecutive epochs of no improvement in validation loss.
Weight decay $10^{-7}$ is used for regularization.

\subsection{Auxiliary Phoneme Loss}
\label{sec:phoneme}

To provide our model with additional training signal and regularize our learned representations, we introduce an auxiliary loss of predicting phoneme labels at each output frame.

To get phoneme labels for each feature frame of the vocalized audio, we use the Montreal Forced Aligner \cite{mcauliffe2017montreal}.
The aligner uses an acoustic model trained on the LibriSpeech dataset in conjunction with a phonemic dictionary to get time-aligned phoneme labels from audio and a transcription.

\begin{table*}[t]
    \centering
    \begin{tabular}{lr}
        \toprule
        \textbf{Model} & \textbf{WER} \\ \midrule
        \citet{gaddy-klein-2020-digital} & 68.0 \\
        This work & \textbf{42.2} \\
        \quad Ablation: Replace convolution features with hand-designed features & 45.2 \\
        \quad Ablation: Replace Transformer with LSTM & 46.0 \\
        \quad Ablation: Remove phoneme loss & 51.7 \\
        \bottomrule
    \end{tabular}
    \caption{Open vocabulary word error rate results from an automatic intelligibility evaluation.}
    \label{tab:results}
\end{table*}

We add an additional linear prediction layer and softmax on top of the Transformer encoder to predict a distribution over phonemes.
For training, we modify the alignment and loss cost $\delta$ by appending a term for phoneme negative log likelihood:
$$\delta'[i,j]=\norm{A_V[i]-\hat{A}_S[j]}_2 - \lambda P_V[i]^\top\log\hat{P}_S[j]$$
where $\hat{P}_S$ is the predicted distribution from the model softmax and $P_V$ is a one-hot vector for the target phoneme label.  We use $\lambda=.1$ for the phoneme loss weight.
After training, the phoneme prediction layer is discarded.

\section{Results}
\label{sec:results}

We train our model on the open-vocabulary data from \citet{gaddy-klein-2020-digital}.
This data contains 19 hours of facial EMG data recordings from a single English speaker during silent and vocalized speech.
Our primary evaluation uses the automatic metric from that work, which transcribes outputs with an automatic speech recognizer\footnote{An implementation of DeepSpeech \cite{hannun2014deep} from Mozilla (\url{https://github.com/mozilla/DeepSpeech})} and compares to a reference with a word error rate (WER) metric.
We also evaluate human intelligibility in Section~\ref{sec:human-eval} below.\footnote{Output audio samples available at \url{https://dgaddy.github.io/silent_speech_samples/ACL2021/}.}

The results of the automatic evaluation are shown in Table~\ref{tab:results}.
Overall, we see that our model improves intelligibility over prior work by an absolute 25.8\%, or 38\% relative error reduction.  Also shown in the table are ablations of our three primary contributions.
We ablate the convolutional feature extraction by replacing those layers with the hand-designed features used in \citet{gaddy-klein-2020-digital}, and we ablate the Transformer layers by replacing with LSTM layers in the same configuration as that work (3 bidirectional layers, 1024 dimensions).
To ablate the phoneme loss, we simply set its weight in the overall loss to zero.
All three of these ablations show an impact on our model's results.

\subsection{Human Evaluation}
\label{sec:human-eval}

In addition to the automatic evaluation, we performed a human intelligibility evaluation using a similar transcription test.  Two human evaluators without prior knowledge of the text were asked to listen to 40 synthesized samples and write down the words they heard (see Appendix~\ref{sec:human-instructions} for full instructions given to evaluators).  We then compared these transcriptions to the ground-truth reference with a WER metric.

The resulting word error rates from the two human evaluators' transcriptions are 36.1\% and 28.5\% (average: 32.3\%), compared to 42.2\% from automatic transcriptions.
These results validate the improvement shown in the automatic metric, and indicate that the automatic metric may be underestimating intelligibility to humans.
However, the large variance across evaluators shows that the automatic metric may still be more appropriate for establishing consistent evaluations across different work on this task.

\section{Phoneme Error Analysis}

One additional advantage to using an auxiliary phoneme prediction task is that
it provides a more easily interpretable view of model predictions.
Although the phoneme predictions are not directly part of the audio synthesis process, we have observed that mistakes in audio and phoneme prediction are often correlated.
Therefore, to better understand the errors that our model makes, we analyze the errors of our model’s phoneme predictions.
To analyze the phoneme predictions, we align predictions on a silent utterance to phoneme labels of a vocalized utterance using the procedure described above in Sections~\ref{sec:training}~and~\ref{sec:phoneme}, then evaluate the phonemes using the measures described in Sections~\ref{sec:confusion}~and~\ref{sec:feature-analysis} below.

\subsection{Confusion}
\label{sec:confusion}

\pgfdeclarelayer{background}
\pgfsetlayers{background,main}
\definecolor{red}{rgb}{.75,0,0}
\let\ipa\textipa
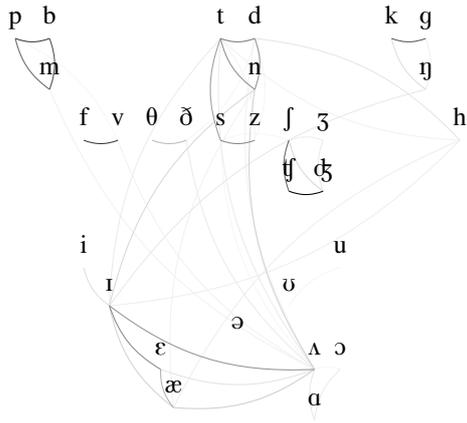
\begin{figure}[t]
    \centering
    \begin{tikzpicture}[x=.5cm,y=-.75cm, scale=.9, every node/.style={scale=.9,font=\vphantom{pb}}]
    \node at (-2, 0) (sil) {};
    
    \begin{scope}[shift={(-7,0)}]
        \node at (1, 0) (p) {p};
        \node at (2, 0) (b) {b};
        \node at (7, 0) (t) {t};
        \node at (8, 0) (d) {d};
        \node at (12, 0) (k) {k};
        \node at (13, 0) (g) {\ipa{g}};
        \node at (2, 1) (m) {m};
        \node at (8, 1) (n) {n};
        \node at (13, 1) (ng) {\ipa{N}};
        \node at (3, 2) (f) {f};
        \node at (4, 2) (v) {v};
        \node at (5, 2) (th) {\ipa{T}};
        \node at (6, 2) (dh) {\ipa{D}};
        \node at (7, 2) (s) {s};
        \node at (8, 2) (z) {z};
        \node at (9, 2) (sh) {\ipa{S}};
        \node at (10, 2) (zh) {\ipa{Z}};
        \node at (14, 2) (hh) {h};
        \node at (9, 3) (ch) {\textteshlig};
        \node at (10,3) (jh) {\textdyoghlig};
    \end{scope}
    
    \begin{scope}[shift={(5,4.5)}, x=.75cm]
    \node at (-1, 0) (uw) {u};
    \node at (-2, .75) (uh) {\ipa{U}};
    \node at (-1, 2) (ao) {\ipa{O}};
    \node at (-1.5, 2) (ah) {\ipa{2}};
    \node at (-1.5, 3) (aa) {\ipa{A}};
    \node at (-3, 1.5) (ax) {\ipa{@}};
    \node at (-6, 0) (iy) {i};
    \node at (-5.5, .75) (ih) {\ipa{I}};
    \node at (-4.5, 2) (eh) {\ipa{E}};
    \node at (-4.25, 2.75) (ae) {\ipa{\ae}};
    \end{scope}
    
\begin{pgfonlayer}{background}
\draw[color=black!3.097975147599511] (sh.south) to[bend right=20] (s.south);  
\draw[color=black!3.1005052854801054] (dh.south) to[bend right=20] (d.south);
\draw[color=black!3.242112920957592] (ao.south) to[bend right=20] (ah.south);
\draw[color=black!3.2860915379579025] (s.south) to[bend right=20] (ah.south);
\draw[color=black!3.2875417130987885] (v.south) to[bend right=20] (p.south);
\draw[color=black!3.344600030460252] (z.south) to[bend right=20] (d.south);   
\draw[color=black!3.5021847040466083] (uw.south) to[bend right=20] (uh.south);
\draw[color=black!4.360849666570816] (z.south) to[bend right=20] (t.south);  
\draw[color=black!4.665000490431523] (t.south) to[bend right=20] (hh.south);
\draw[color=black!5.032332664523499] (n.south) to[bend right=20] (ae.south);
\draw[color=black!5.053706907002658] (v.south) to[bend right=20] (ah.south);
\draw[color=black!5.08990615187601] (m.south) to[bend right=20] (ah.south);
\draw[color=black!5.158283438739451] (d.south) to[bend right=20] (ah.south);
\draw[color=black!5.779036143837281] (ng.south) to[bend right=20] (g.south);
\draw[color=black!6.212707025766311] (ih.south) to[bend right=20] (hh.south);
\draw[color=black!6.3759681697680985] (hh.south) to[bend right=20] (ae.south);
\draw[color=black!6.546497141752276] (zh.south) to[bend right=20] (sh.south);
\draw[color=black!7.257900360624199] (dh.south) to[bend right=20] (ah.south);
\draw[color=black!7.564245854079567] (t.south) to[bend right=20] (ih.south);
\draw[color=black!7.571082351983911] (t.south) to[bend right=20] (ah.south);
\draw[color=black!7.737009698900474] (zh.south) to[bend right=20] (jh.south);
\draw[color=black!7.844212928754492] (ao.south) to[bend right=20] (aa.south);
\draw[color=black!8.675105498382361] (ng.south) to[bend right=20] (ih.south);
\draw[color=black!8.937938917532048] (ah.south) to[bend right=20] (aa.south);
\draw[color=black!9.916569190213659] (hh.south) to[bend right=20] (d.south);
\draw[color=black!10.433442137671886] (eh.south) to[bend right=20] (ah.south);
\draw[color=black!12.07350447439989] (iy.south) to[bend right=20] (ih.south);
\draw[color=black!12.709121191598951] (ng.south) to[bend left=20] (k.south);
\draw[color=black!14.748763532102146] (sh.south) to[bend right=20] (jh.south);
\draw[color=black!15.09475059928192] (n.south) to[bend right=20] (ih.south);
\draw[color=black!17.19178349770413] (n.south) to[bend right=20] (ah.south);
\draw[color=black!17.39742092779357] (ih.south) to[bend right=20] (ae.south);
\draw[color=black!18.51566553309018] (ah.south) to[bend left=20] (ae.south);
\draw[color=black!25.872198919215993] (th.south) to[bend right=20] (dh.south);
\draw[color=black!30.631359152328585] (eh.south) to[bend right=20] (ae.south);
\draw[color=black!37.574416217014814] (t.south) to[bend right=20] (s.south);
\draw[color=black!41.2125672741966] (ih.south) to[bend right=20] (ah.south);
\draw[color=black!41.262430954026556] (n.south) to[bend right=20] (d.south);
\draw[color=black!43.46850498061906] (t.south) to[bend right=20] (n.south);
\draw[color=black!45.497733278778156] (ih.south) to[bend right=20] (eh.south);
\draw[color=black!46.3595253282884] (z.south) to[bend left=20] (s.south);
\draw[color=black!50.95530217871406] (t.south) to[bend right=20] (d.south);
\draw[color=black!58.74137243901648] (p.south) to[bend right=20] (m.south);
\draw[color=black!60.01903669170171] (sh.south) to[bend right=20] (ch.south);
\draw[color=black!64.87625343252574] (k.south) to[bend right=20] (g.south);
\draw[color=black!68.14406250696229] (m.south) to[bend right=20] (b.south);
\draw[color=black!76.28755538333134] (p.south) to[bend right=20] (b.south);
\draw[color=black!76.84391705800043] (v.south) to[bend left=20] (f.south);
\draw[color=black!100.0] (jh.south) to[bend left=20] (ch.south);

\end{pgfonlayer}
    
    \end{tikzpicture}
    \caption{Phoneme confusability (darker lines indicate more confusion - maximum darkness is 13\% confusion)}
    \label{fig:confusion}
\end{figure}

First, we measure the confusion between each pair of phonemes.  We use a frequency-normalized metric for confusion: $(e_{p1,p2}+e_{p2,p1})/(f_{p1}+f_{p2})$,
where $e_{p1,p2}$ is the number of times $p2$ was predicted when the label was $p1$, and $f_{p1}$ is the number of times phoneme $p1$ appears as a target label.
Figure~\ref{fig:confusion} illustrates this measure of confusion using darkness of lines between the phonemes, and Appendix~\ref{sec:phoneme-chart} lists the values of the most confused pairs.

We observe that many of the confusions are between pairs of consonants that differ only in voicing, which is consistent with the observation in \citet{gaddy-klein-2020-digital} that voicing signals appear to be subdued in silent speech.
Another finding is a confusion between nasals and stops, which is challenging due to the role of the velum and its relatively large distance from the surface electrodes, as has been noted in prior work \cite{freitas2014velum}.
We also see some confusion between vowel pairs and between vowels and consonants, though these patterns tend to be less interpretable. 

\subsection{Articulatory Feature Accuracy}
\label{sec:feature-analysis}

To better understand our model's accuracy across different consonant articulatory features, we perform an additional analysis of phoneme selection across specific feature dimensions.
For this analysis, we define a confusion set for an articulatory feature as a set of English phonemes that are identical across all other features.
For example, one of the confusion sets for the place feature is \emph{\{p,~t,~k\}}, since these phonemes differ in place of articulation but are the same along other axes like manner and voicing (a full listing of confusion sets can be found in Appendix~\ref{sec:feature-analysis-details}).
For each feature of interest, we calculate a forced-choice accuracy within the confusion sets for that feature.
More specifically,
we find all time steps in the target sequence with labels belonging in a confusion set and restrict our model output to be within the corresponding set for those positions.
We then compute an accuracy across all those positions that have a confusion set.

To evaluate how much of the articulatory feature accuracies can be attributed to contextual inferences rather than information extracted from EMG, we compare our results to a baseline model that is trained to make decisions for a feature based on nearby phonemes.
In the place of EMG feature inputs, this baseline model is given the sequence of phonemes predicted by the full model, but with information about the specific feature being tested removed by collapsing phonemes in each of its confusion sets to a single symbol.
Additional details on this baseline model can be found in Appendix~\ref{sec:feature-analysis-details}.

The results of this analysis are shown in Figure~\ref{fig:feature-analysis}.
By comparing the gap in accuracy between the full model and the phoneme context baseline, we again observe trends that correspond to our prior expectations.  While place and oral manner features can be predicted much better by our EMG model than from phonemic context alone, nasality and voicing are more challenging and have a smaller improvement over the contextual baseline.

\definecolor{bblue}{HTML}{4F81BD}
\definecolor{rred}{HTML}{C0504D}
\definecolor{ggrey}{HTML}{9F9F9F}
\definecolor{ggreen}{HTML}{9BBB59}
\definecolor{ppurple}{HTML}{9F4C7C}

\begin{figure}[t]
    \centering

\begin{tikzpicture}
    \begin{axis}[
        width  = .91\columnwidth,
        height = 6.5cm,
        major y tick style = transparent,
        xbar=2*\pgflinewidth,
        bar width=8pt,
        xmajorgrids = true,
        xlabel = {Accuracy},
        symbolic y coords={Place,Oral manner,Nasality,Voicing},
        ytick = data,
        scaled x ticks = false,
        enlarge y limits=0.2,
        xmin=35,
        xmax=100,
        y dir=reverse, 
        reverse legend,
        legend cell align=left,
        legend style={
                at={(1.06,1.01)}, 
                anchor=south east,
                column sep=.2ex,
                /tikz/every even column/.append style={column sep=1.2ex},
                draw=none,
                legend columns=3,
                font=\small
        },
        label style={font=\small},
        tick label style={font=\small},
        legend image code/.code={%
            \draw[#1, draw=none] (0cm,-0.1cm) rectangle (0.3cm,0.15cm);
        },
        x label style={at={(axis description cs:0.5,0.03)}}
    ]
        \addplot[style={ggrey,fill=ggrey,mark=none}]
             coordinates {(56.2,Place) (40.5,Oral manner) (64.5,Nasality) (65.1,Voicing)};
        
        \addplot[style={rred,fill=rred,mark=none}]
            coordinates {(80.4,Place) (79.6,Oral manner) (84.3,Nasality) (85.3,Voicing)};

        \addplot[style={bblue,fill=bblue,mark=none}]
             coordinates {(96.6,Place) (92.9,Oral manner) (89.1,Nasality) (88.1,Voicing)};

        \legend{Majority class,Phoneme context,Full context}
    \end{axis}
\end{tikzpicture}

    \caption{Accuracy of selecting phonemes along articulatory feature dimensions.  We compare our full EMG model (full context) with a majority class baseline and a model given only phoneme context as input.}
    \label{fig:feature-analysis}
\end{figure}
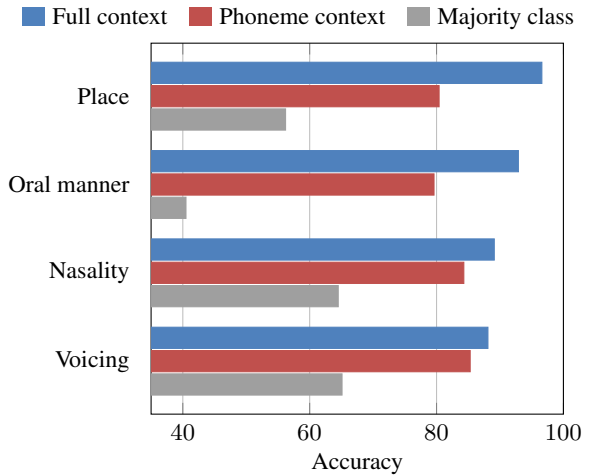

\section{Conclusion}

By improving several model components for voicing silent speech, our work has achieved a 38\% relative error reduction on this task.
Although the problem is still far from solved, we believe the large rate of improvement is a promising sign for continued progress.

\section*{Acknowledgments}

This  material  is  based  upon  work  supported  by the National Science Foundation under Grant No. 1618460 and by DARPA under the LwLL program / Grant No. FA8750-19-1-0504.

\bibliographystyle{acl_natbib}
\bibliography{anthology,acl2021}

\appendix

\section{Instructions to Human Evaluators}
\label{sec:human-instructions}

The following instructions were given to human evaluators for the transcription test described in Section~\ref{sec:human-eval}:

\emph{Please listen to each of the attached sound files and write down what you hear.  There are 40 files, each of which will contain a sentence in English.  Write your transcriptions into a spreadsheet such as Excel or Google sheets so that the row numbers match the numbers in the file names.  Many of the clips may be difficult to hear.  If this is the case, write whatever words you are able to make out, even if it does not form a complete expression.  If you are not entirely sure about a word but can make a strong guess, you may include it in your transcription, but only do so if you beleive it is more likely than not to be the correct word.  If you cannot make out any words, leave the corresponding row blank.}

\section{Phoneme Confusability}
\label{sec:phoneme-chart}

This section provides numerical results for phoneme confusions to complement the illustration given in Section~\ref{sec:confusion} of the main paper.  We compare the frequency of errors between two phonemes to the frequency of correct predictions on those phonemes.  We define the following two quantities:
$$\text{Confusion: }(e_{p1,p2}+e_{p2,p1})/(f_{p1}+f_{p2})$$
$$\text{Accuracy: }(e_{p1,p1}+e_{p2,p2})/(f_{p1}+f_{p2})$$
where $e_{p1,p2}$ is the number of times $p2$ was predicted when the label was $p1$, and $f_{p1}$ is the number of times phoneme $p1$ appears as a target label.  Results for the most confused pairs are shown in the table below.

\begin{center}
\begin{tabular}{c c c c}
\toprule
\multicolumn{2}{c}{Phonemes} & Confusion (\%) & Accuracy (\%) \\ \midrule
\textdyoghlig & \textteshlig & 13.2 & 49.4 \\
v & f & 10.4 & 72.0 \\
p & b & 10.3 & 64.3 \\
m & b & 9.3 & 74.3 \\  
k & \ipa{g} & 8.9 & 77.2 \\ 
\ipa{S} & \textteshlig & 8.3 & 59.8 \\
p & m & 8.1 & 73.0 \\  
t & d & 7.2 & 64.0 \\
z & s & 6.6 & 80.0 \\  
\ipa{I} & \ipa{E} & 6.5 & 60.6 \\
t & n & 6.3 & 67.1 \\ 
n & d & 6.0 & 66.8 \\ 
\ipa{I} & \ipa{2} & 6.0 & 65.8 \\                    
\ipa{\*r} & \textrhookschwa & 5.7 & 78.2 \\
t & s & 5.5 & 72.8 \\
\ipa{E} & \ipa{\ae} & 4.7 & 70.9 \\
u & o\ipa{U} & 4.3 & 77.4 \\
\ipa{T} & \ipa{D} & 4.1 & 76.9 \\
\ipa{2} & \ipa{\ae} & 3.2 & 72.1 \\
\ipa{I} & \ipa{\ae} & 3.1 & 64.9 \\
\bottomrule
\end{tabular}
\end{center}

\section{Articulatory Feature Analysis Details}
\label{sec:feature-analysis-details}

The following table lists all confusion sets used in our articulatory feature analysis in Section~\ref{sec:feature-analysis}.

\begin{center}
\begin{tabular}{l c}
    \toprule
    Feature & Confusion Sets  \\
    \midrule
    Place & \{p,t,k\} \{b,d,g\} \{m,n,\ipa{N}\} \\
    &\{f,\ipa{T},s,\ipa{S},h\} \{v,\ipa{D},z,\ipa{Z}\} \\
    Oral manner & \{t,s\} \{d,z,l,r\} \{\ipa{S},\textteshlig\} \{\ipa{Z},\textdyoghlig\} \\
    Nasality & \{b,m\} \{d,n\} \{g,\ipa{N}\} \\
    Voicing & \{p,b\} \{t,d\} \{k,g\} \{f,v\} \\
    & \{\ipa{T},\ipa{D}\} \{s,z\} \{\ipa{S},\ipa{Z}\} \{\textteshlig,\textdyoghlig\} \\
    \bottomrule
\end{tabular}
\end{center}

The phoneme context baseline model uses a Transformer architecture with dimensions identical to our primary EMG-based model, but is fed phoneme embeddings of dimension 768 in the place of the convolutional EMG features.  The phonemes input to this model are the maximum-probability predictions output by our primary model at each frame, but with all phonemes from a confusion set replaced with the same symbol.  We train a separate baseline model for each of the four articulatory feature types to account for different collapsed sets in the input.  During training, a phoneme likelihood loss is applied to all positions and no restrictions are enforced on the output.  Other training hyperparameters are the same between this baseline and the main model.

\section{Additional Reproducability Information}
All experiments were run on a single Quadro RTX 6000 GPU, and each took approximately 12 hours.  Hyperparameters were tuned manually based on automatic transcription WER on the validation set.  The phoneme loss weight hyperparameter $\lambda$ was chosen from $\{1,.5,.1,.05,.01,.005\}$.  We report numbers on the same test split as \citet{gaddy-klein-2020-digital}, but increase the size of the validation set to 200 examples to decrease variance during model exploration and tuning.  Our model contains approximately 40 million parameters.

\end{document}